\documentclass[12pt,english]{amsart}
\usepackage[T1]{fontenc}
\usepackage[latin9]{inputenc}
\usepackage{geometry}
\usepackage{babel}
\usepackage{amsbsy}
\usepackage{amstext}
\usepackage{amsthm}
\usepackage{amssymb}
\usepackage{graphicx}
\usepackage[unicode=true,pdfusetitle,
 bookmarks=true,bookmarksnumbered=false,bookmarksopen=false,
 breaklinks=true,pdfborder={0 0 1},backref=false,colorlinks=false]
 {hyperref}

\makeatletter
\numberwithin{equation}{section}
\numberwithin{figure}{section}
\theoremstyle{plain}
\newtheorem{thm}{\protect\theoremname}[section]
\theoremstyle{plain}
\newtheorem{prop}[thm]{\protect\propositionname}
\theoremstyle{remark}
\newtheorem{rem}[thm]{\protect\remarkname}
\theoremstyle{plain}
\newtheorem{cor}[thm]{\protect\corollaryname}
\theoremstyle{remark}
\newtheorem*{acknowledgement*}{\protect\acknowledgementname}

\usepackage{xurl}

\newcommand*{\rom}[1]{\expandafter\@slowromancap\romannumeral #1@}
\newcommand*{\ara}[1]{\expandafter\@slowromancap\arabic #1@}

\makeatletter
\def\blfootnote{\xdef\@thefnmark{}\@footnotetext}
\makeatother

\makeatother

\providecommand{\acknowledgementname}{Acknowledgement}
\providecommand{\corollaryname}{Corollary}
\providecommand{\propositionname}{Proposition}
\providecommand{\remarkname}{Remark}
\providecommand{\theoremname}{Theorem}

\begin{document}
\global\long\def\B{\mathcal{B}}%

\global\long\def\R{\mathbb{R}}%

\global\long\def\Q{\mathbb{Q}}%

\global\long\def\Z{\mathbb{Z}}%

\global\long\def\C{\mathbb{C}}%

\global\long\def\N{\mathbb{N}}%

\global\long\def\RR{\mathbb{\overline{R}}}%

\global\long\def\rme{\mathrm{e}}%

\global\long\def\rmi{\mathrm{i}}%

\global\long\def\rmd{\mathrm{d}}%

\global\long\def\E{\mathcal{E}}%

\global\long\def\V{\mathcal{V}}%

\global\long\def\P{\mathcal{P}}%

\global\long\def\VR{\mathcal{V}_{\mathcal{R}}}%

\global\long\def\dom{\Omega}%

\global\long\def\sdom{\mathcal{S}}%

\global\long\def\spec#1{\textrm{\text{Spec}}\left(#1\right)}%

\global\long\def\meig#1{\mathcal{\lambda}^{\Gamma}\left(#1\right)}%

\global\long\def\deig#1{\lambda\left(#1\right)}%

\global\long\def\sreg{\Sigma^{\text{reg}}}%

\global\long\def\sf#1{\mathrm{Sf}_{#1}}%

\global\long\def\mult#1#2{\mathrm{Mult}_{#1}\left(#2\right)}%

\global\long\def\L{L}%

\global\long\def\C{\mathbb{C}}%

\global\long\def\R{\mathbb{R}}%

\global\long\def\N{\mathbb{N}}%

\global\long\def\V{\mathcal{V}}%

\global\long\def\E{\mathcal{E}}%

\global\long\def\VR{\mathcal{V_{R}}}%

\global\long\def\Id{\mathbb{\boldsymbol{I}}}%

\global\long\def\Bell{\mathbb{\ensuremath{\boldsymbol{\ell}}}}%

\global\long\def\nmean#1{\langle#1\rangle_{n}}%

\global\long\def\hmean#1{\langle#1\rangle_{h}}%

\global\long\def\sreg{\Sigma^{\text{reg}}}%

\global\long\def\deltak{\langle\Delta k\rangle}%

\global\long\def\ncf{\mathcal{N}}%

\global\long\def\lmin{l_{min}}%

\global\long\def\Uv{\mathcal{U}_{v}}%

\global\long\def\Sv{\mathcal{S}_{v}}%

\global\long\def\tr{\text{tr}}%

\global\long\def\Ev{\mathcal{E}_{v}}%

\title[The Schrödinger equation on time dependent quantum graphs]{Time evolution and the Schrödinger equation on time dependent quantum
graphs}
\author{Uzy Smilansky \and Gilad Sofer}
\dedicatory{Dedicated to Michael Berry on his forthcoming anniversary.}
\begin{abstract}
The purpose of the present paper is to discuss the time dependent
Schrödinger equation on a metric graph with time-dependent edge lengths,
and the proper way to pose the problem so that the corresponding time
evolution is unitary. We show that the well posedness of the Schrödinger
equation can be guaranteed by replacing the standard Kirchhoff Laplacian
with a magnetic Schrödinger operator with a harmonic potential. We
then generalize the result to time dependent families of vertex conditions.
We also apply the theory to show the existence of a geometric phase
associated with a slowly changing quantum graph.
\end{abstract}

\maketitle
\blfootnote{Uzy Smilansky,
Department of Physics of Complex Systems, Weizmann Institute of Science, Rehovot 7610001, Israel.
Email: uzy.smilansky@weizmann.ac.il

Gilad Sofer,
Faculty of Mathematics, Technion - Israel Institute of Technology, Haifa 32000, Israel.
Email: gilad.sofer@campus.technion.ac.il}

\section{Introduction}

\label{section1}

In the present work, we study the time evolution on quantum graphs
whose edge lengths vary in time. A stationary quantum graph consists
of a set of vertices connected by edges of prescribed lengths. On
each edge, the Schrödinger operator is the one-dimensional Laplacian.
The graph is supplemented with appropriate boundary conditions at
the vertices, which ensure that the operator is self-adjoint and so
the resulting time evolution is unitary. Our goal is to generalize
this model by allowing the edge lengths to be time dependent.

To illustrate the possible difficulties this time dependence causes,
we start by considering the simplest example. Namely, a graph that
consists of two vertices connected by an edge of length $L(t/T)$.
Here, $T^{-1}$ measures the rate of change of the edge lengths.

The evolution for this system is dictated by the time-dependent Schrödinger
equation: 
\begin{eqnarray}
 &  & i\hbar\frac{\partial\psi\left(x,t\right)}{\partial t}=-\frac{\hbar^{2}}{2m}\frac{\partial^{2}\psi\left(x,t\right)}{\partial x^{2}}\ \ ,\ \ \ x\in\left[-\frac{L\left(t/T\right)}{2},\frac{L\left(t/T\right)}{2}\right].\label{td interval}
\end{eqnarray}
At the two vertices we impose the standard Neumann boundary condition:
\begin{eqnarray}
\frac{\partial\psi\left(x,t\right)}{\partial x}=0\ \ \ {\rm at\ }\ \left|x\right|=\frac{L\left(t/T\right)}{2},\ \forall t\geq0.
\end{eqnarray}
The operator on the right hand side of (\ref{td interval}) involves
only derivatives with respect to $x$. The time dependence is due
to the boundary conditions, which are imposed at the time dependent
boundary. If one considers the time $t$ as a parameter, the operator
on the right is referred to as the \textit{instantaneous Hamiltonian}.

\noindent Given some initial condition $\psi\left(x,0\right)$, we
let the system evolve according to (\ref{td interval}). A straightforward
computation then gives the following expression for the time derivative
of the squared $L^{2}$ norm of $\psi\left(x,t\right)$: 
\begin{align}
 & \frac{{\rm d}\ }{{\rm d}t}\Vert\psi\Vert^{2}=\frac{{\rm d}\ }{{\rm d}t}\int_{-\frac{L(t/T)}{2}}^{\frac{L(t/T)}{2}}|\psi\left(x,t\right)|^{2}{\rm d}x\label{normchange}\\
 & =\frac{1}{2T}\cdot\frac{{\rm d}L\left(t/T\right)}{{\rm d}t}\left[\left|\psi\left(\frac{L\left(t/T\right)}{2},t\right)\right|^{2}+\left|\psi\left(-\frac{L\left(t/T\right)}{2},t\right)\right|^{2}\right],\nonumber 
\end{align}
This rather unexpected change of the $L^{2}$ norm in time is due
to the choice of Neumann boundary condition, which causes the additional
term $\int_{-\frac{L(t/T)}{2}}^{\frac{L(t/T)}{2}}\frac{{\rm d}\ }{{\rm d}t}|\psi\left(x,t\right)|^{2}{\rm d}x$
to vanish after integrating by parts.

\noindent Recalling the standard interpretation of $|\psi(x,t)|^{2}dx$
as the probability for finding the particle in the interval $\left(x,x+dx\right)$
at time $t$, the above result implies that there exists a probability
flux which equals to the product of the density at the boundary and
its speed. Hence, the probability to find the particle within the
boundaries changes in time. This implies that the evolution dictated
by (\ref{td interval}) is not unitary. In this context, it is worth
noting:

\noindent 1. The instantaneous $L^{2}$ norm is constant in the adiabatic
limit $T\rightarrow\infty$.

\noindent 2. For the Dirichlet boundary condition ($\psi(x,t)=0$),
the instantaneous $L^{2}$ norm is constant. Equation (\ref{normchange})
shows that the change in norm arises also for the Robin boundary condition:
\begin{equation}
\frac{\partial\psi}{\partial x}\left(x,t\right)=C\psi\left(x,t\right)\ \ \ {\rm at\ }\ \left|x\right|=\frac{L\left(t/T\right)}{2},\,C\in\mathbb{R}.
\end{equation}

In Section \ref{section2} we return to this problem, and derive the
boundary conditions which render the time evolution unitary, ensuring
probability conservation. We follow previous studies which treat time
dependent domains in different contexts. The time dependent Schrödinger
equation on domains in $\mathbb{R}^{n}$ with time dependent boundary
was studied in \cite{Doescher1969,Munier1981,Duca2021}. The propagation
of electromagnetic waves in vibrating cavities is a closely related
problem studied in \cite{Moore1970}. The present work can be considered
as an extension of these studies to quantum graphs. We shall note
that the time dependent Schrödinger equation on a star graph was studied
in \cite{Matrasulov2015}. However, the vertex boundary conditions
considered there correspond to the Dirichlet condition, and thus do
not address the problem discussed above.

Once the problem is properly reformulated in Section \ref{section2},
the effect of the moving boundary is expressed in terms of a gauge
field which, when included, renders the Schrödinger operator self-adjoint
(Theorem \ref{theorem-kirchhoff}). This is analogous to the derivation
for slowly varying time dependent domains presented in \cite{Duca2021}. 

In Section \ref{sec:Generalizations}, we generalize the result for
graphs where the vertex conditions depend on time as well (Theorem
\ref{thm:general-condition}). Moreover, we apply these results by
studying the geometric phase associated with the time evolution in
the adiabatic limit. Finally, in Appendix \ref{sec:Secular} we derive
the secular equation whose zeros give the instantaneous spectrum at
each time.

\bigskip

\section{Quantum graphs with time dependent edge lengths and standard boundary
conditions\label{section2}}

This section consists of three parts, in which the time dependent
Schrödinger equation for quantum graphs with time dependent edge lengths
is formulated and discussed. The first subsection addresses the problem
in the simple case of a single interval. It serves as a primer for
the treatment of a general graph which is presented in the second
subsection. In the third subsection we demonstrate the theory by studying
the time evolution on a time dependent equilateral quantum graph.

\subsection{A time dependent interval\label{interval}}

Before proceeding further, it is worthwhile to rewrite the Schrödinger
equation (\ref{td interval}) in dimensionless form. Observing that
$\frac{\hbar}{2m}$ has the dimension of $\frac{\text{length}^{2}}{\text{time}}$,
then scaling $x$ by $L$ and $t$ by $T$ results in the dimensionless
quantity $\frac{L^{2}}{T}\cdot\frac{\hbar}{2m}$. Using the standard
convention in which $\frac{\hbar}{2m}$ takes the numerical value
$1$, we remain with a dimensionless $\frac{L^{2}}{T}\cdot\frac{\hbar}{2m}$.
We then get the following Schrödinger equation: 
\begin{eqnarray}
 &  & i\frac{\partial\psi\left(x,t\right)}{\partial t}=-\frac{\partial^{2}\psi\left(x,t\right)}{\partial x^{2}}\ \ ,\ \ \ x\in\left[-\frac{1}{2}L\left(t/T\right),\frac{1}{2}L\left(t/T\right)\right].\label{td2interval}
\end{eqnarray}

Following \cite{Doescher1969,Munier1981,Duca2021}, we wish to write
the Schrödinger equation in terms of the scaled variables 
\begin{equation}
\tau=\frac{t}{T}\ ,\ \xi=\frac{x}{L\left(t/T\right)}.\label{coordinates}
\end{equation}
To do this, we introduce the unitary transformation: 
\begin{eqnarray}
 &  & \omega\left(\xi,\tau\right):=L\left(\tau\right)^{1/2}\psi\left(L\left(\tau\right)\xi,T\tau\right),\label{unitary}
\end{eqnarray}
where now $\tau\in\left[0,1\right]$ and $\xi\in\left[-\frac{1}{2},\frac{1}{2}\right]$.
The Schrödinger equation (\ref{td2interval}) can be rewritten in
the new coordinates as 
\begin{eqnarray}
 &  & \frac{i}{T}\frac{\partial}{\partial\tau}\omega\left(\xi,\tau\right)\label{version1}\\
 &  & =\frac{1}{L\left(\tau\right)^{2}}\left[-\left(\frac{\partial}{\partial\xi}-i\xi\frac{\dot{L}\left(\tau\right)L\left(\tau\right)}{2T}\right)^{2}-\xi^{2}\left(\frac{\dot{L}\left(\tau\right)L\left(\tau\right)}{2T}\right)^{2}\right]\omega\left(\xi,\tau\right).\nonumber 
\end{eqnarray}

In the new Schrödinger equation, the Schrödinger operator includes
a repelling harmonic potential, and the Laplacian appears as a magnetic
Laplacian with $\mathcal{A}\left(\xi,\tau\right)=\xi\frac{\dot{L}(\tau)L(\tau)}{2T}$
playing the role of a vector potential. This suggests that one can
render the magnetic operator above self-adjoint by replacing the original
Neumann condition with the magnetic Neumann condition: 
\begin{equation}
\left(\frac{\partial}{\partial\xi}-i\xi\frac{\dot{L}L}{2T}\right)\omega\left(\xi,\tau\right)=0\ \ {\rm at\ }\ |\xi|=\frac{1}{2}.\label{magnetic1}
\end{equation}
 This indeed gives a self-adjoint operator on $L^{2}\left(-\frac{1}{2},\frac{1}{2}\right)$.
Using (\ref{magnetic1}) and (\ref{unitary}), we obtain the corresponding
boundary condition for $\psi\left(x,t\right)$: 
\begin{equation}
\frac{\partial\psi\left(x,t\right)}{\partial x}=\pm\frac{i\dot{L}}{4T}\psi\left(x,t\right)\ \ {\rm at\ }\ x=\pm\frac{1}{2}L\left(t/T\right).\label{Robinmod}
\end{equation}
 Repeating the computation in (\ref{normchange}) with the new boundary
condition immediately shows that the resulting time evolution is unitary.

The following alternative explanation for the magnetic boundary condition
was suggested by Michael Berry \cite{Berry}. The movement of the
edges induces a nontrivial probability current at the boundary. One
may account for this probability current by introducing a \emph{modified}
Robin boundary condition: 
\begin{equation}
\frac{\partial\psi\left(x,t\right)}{\partial x}=C\psi\left(x,t\right)\ \ {\rm at\ }\ |x|=\frac{1}{2}L\left(t/T\right),\,\,C\in\mathbb{C}.
\end{equation}
 While the usual Robin condition corresponds to $C\in\mathbb{R}$,
a straightforward computation shows that the requirement $\frac{{\rm d}}{{\rm d}t}\Vert\psi\Vert^{2}=0$
gives the condition $Im\left(C\right)=\frac{\dot{L}}{4T}$. We may
thus solve the problem by suggesting the boundary condition (\ref{Robinmod}).
The associated Robin parameter may be considered as an effective magnetic
field.

The Schrödinger equation and boundary conditions in (\ref{version1}),
(\ref{magnetic1}) can be further simplified by introducing the gauge
transformation 
\begin{equation}
\omega\left(\xi,\tau\right)=e^{\frac{i}{2}\theta\left(\tau\right)\left(\xi^{2}-1/4\right)}g\left(\xi,\tau\right).
\end{equation}
 One may then eliminate the magnetic Laplacian in Equation (\ref{version1})
if the phase is chosen to be $\theta\left(\tau\right)=\frac{\dot{L}L}{2T}$.
Note that in the units used in this work, $\theta$ is dimensionless.

The Schrödinger equation is now given by 
\begin{align}
 & \frac{i}{T}\frac{\partial g\left(\xi,\tau\right)}{\partial\tau}=\label{final}\\
 & -\frac{1}{L\left(\tau\right)^{2}}\cdot\frac{\partial^{2}g\left(\xi,\tau\right)}{\partial\xi^{2}}+\frac{\ddot{L}L}{4T^{2}}\xi^{2}g\left(\xi,\tau\right)-\frac{\ddot{L}L\left(\tau\right)+\dot{L}^{2}\left(\tau\right)}{16T^{2}}g\left(\xi,\tau\right),\ |\xi|\leq\frac{1}{2},\nonumber 
\end{align}
and subject to standard Neumann boundary conditions.

A similar equation was presented in \cite{Matrasulov2015,Munier1981}.
We derived it here as a prelude to the treatment of the time dependent
Schrödinger operator on a graph. Note that the factor $\frac{1}{T}$
multiplies the $\tau$ derivative, implying that the adiabatic limit
$T\rightarrow\infty$ is the analogue of the semi-classical limit.

\subsection{Compact metric graphs with time dependent edge lengths\label{graph}}

Consider now the metric graph $\Gamma=\left(\mathcal{V},\mathcal{E},\mathcal{L}\right)$
, where $\mathcal{V}$ is the vertex set and $\mathcal{E}$ is the
edge set (both assumed finite). Moreover, $\mathcal{L}:\left[0,T\right]\rightarrow\mathbb{R}^{\left|\mathcal{E}\right|}$
is a family of edge lengths parameterized as $\left\{ L_{e}\left(t/T\right)\right\} _{e\in\mathcal{E}}$
with $L_{e}\left(t/T\right)$ positive and twice continuously differentiable.
A coordinate $x_{e}$ with $x_{e}\in\left[-\frac{1}{2}L_{e}(t/T),\frac{1}{2}L_{e}(t/T)\right]$
is assigned to every edge in $\mathcal{E}$. We denote the set of
edges connected to the vertex $v$ by $S_{v}$ and the degree of the
vertex by $d_{v}:=|S_{v}|$.

\begin{figure}
\centerline{ \includegraphics[width=0.45\textwidth]{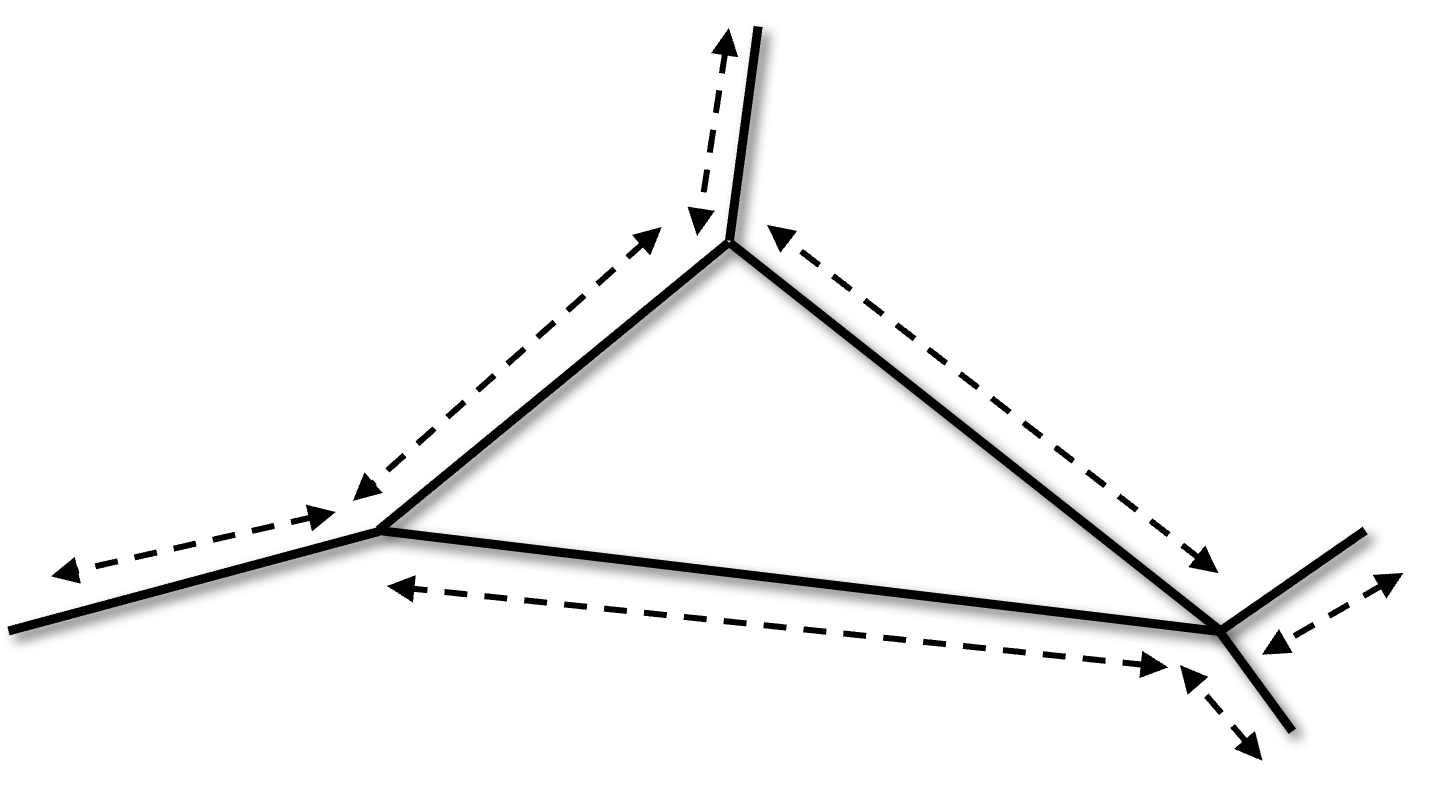}
} \caption{\label{fig:t-graph} A time dependent metric graph.}
\end{figure}

The time dependent Schrödinger operator for this model consists of
the direct sum of $\left|\mathcal{E}\right|$ one-dimensional Laplacians
attached to the edges ${\bf \Delta}=\bigoplus_{e\in\mathcal{E}}\frac{\partial^{2}}{\partial x_{e}^{2}}$:
\begin{equation}
i\frac{\partial}{\partial t}\Psi\left[{\bf x},t\right]=-{\bf \Delta}\Psi\left[{\bf x},t\right],\ \ \ \ \ {\bf x}=\{x_{e}\}_{e\in\mathcal{E}},\ \label{laplacian}
\end{equation}
 with 
\begin{equation}
\Psi\left[{\bf x},t\right]=\left\{ \psi_{e}\left(x_{e},t\right)\right\} _{e\in\mathcal{E}}\ ,\ \ \ 0\le t\le T.
\end{equation}
 We have shown in the preceding section that for the graph which consists
of a single interval, the standard Neumann boundary conditions that
are used in the time independent case should be modified when the
length is time dependent. A similar computation shows that the same
is true for arbitrary graphs. Namely, one should replace the commonly
used Neumann-Kirchhoff vertex condition: 
\begin{eqnarray}
\hspace{-10mm} &  & \psi_{e}\left(x_{e}=v,t\right)=\psi_{e'}\left(x_{e'}=v,t\right),\ \ \ \ \forall e,e'\in S_{v},\label{NK}\\
\hspace{-10mm} &  & \sum_{e\in S_{v}}\frac{\partial\psi_{e}\left(v,t\right)}{\partial x_{e}}=0,\ \ ({\rm derivatives\ directed\ away\ from\ the\ vertex}).\nonumber 
\end{eqnarray}
 The alternative formulation should provide a self-adjoint instantaneous
Schrödinger operator, which, in the limit of constant edge lengths,
should converge to the standard Neumann-Kirchhoff Laplacian in an
appropriate sense. We focus on the standard vertex conditions (\ref{NK})
in this section; in Section \ref{sec:Generalizations} we provide
a treatment of more general vertex conditions.

Following the example of the interval, it is suggested that the proper
way to represent the problem is by introducing a new set of coordinates,
\begin{equation}
\xi_{e}=\frac{x_{e}}{L_{e}\left(t/T\right)},\tau=\frac{t}{T},\label{eq:graph-coordinates}
\end{equation}
 in which the edge lengths are identically equal to one. Denoting
the associated (stationary) metric graph by $\Gamma_{0}$, we define
the family of unitaries 
\begin{eqnarray}
 &  & U_{t}:L^{2}\left(\Gamma\right)\rightarrow L^{2}\left(\Gamma_{0}\right),\label{unitary3}\\
 &  & U_{t}\psi_{e}\left(x_{e},t\right)=\omega_{e}\left(\xi_{e},\tau\right):=L_{e}\left(\tau\right)^{1/2}\psi\left(L_{e}\left(\tau\right)\xi_{e},T\tau\right).\label{unitary4}
\end{eqnarray}
 As before, the family $U_{t}$ gives us the new Schrödinger equation
on each edge:
\begin{align}
 & \frac{i}{T}\frac{\partial}{\partial\tau}\omega_{e}\left(\xi_{e},\tau\right)=\label{grapheq}\\
 & \frac{1}{L_{e}\left(\tau\right)^{2}}\left[-\left(\frac{\partial}{\partial\xi_{e}}-i\xi_{e}\frac{\dot{L}_{e}\left(\tau\right)L_{e}\left(\tau\right)}{2T}\right)^{2}-\xi_{e}^{2}\left(\frac{\dot{L}_{e}\left(\tau\right)L_{e}\left(\tau\right)}{2T}\right)^{2}\right]\omega_{e}\left(\xi_{e},\tau\right).\nonumber 
\end{align}

The vertex boundary conditions which correspond to the magnetic Schrödinger
operator above are obtained by precomposing the vertex condition (\ref{NK})
with $U_{t}^{-1}$, and replacing $\frac{\partial}{\partial\xi}$
with the magnetic derivative associated with (\ref{grapheq}). This
gives 
\begin{eqnarray}
 &  & L_{e}^{-1/2}\left(\tau\right)\omega_{e}\left(v,\tau\right)=L_{e'}^{-1/2}\left(\tau\right)\omega_{e'}\left(v,\tau\right),\ \ \forall e,e'\in S_{v},\label{graphvc}\\
 &  & \sum_{e\in S_{v}}\frac{1}{L_{e}^{3/2}\left(\tau\right)}\left(\frac{\partial}{\partial\xi_{e}}-i\xi_{e}\frac{\dot{L}_{e}L_{e}}{2T}\right)\omega_{e}\left(\xi_{e},\tau\right)=0\ \ {\rm at\ }\ |\xi_{e}|=\frac{1}{2}.\label{graphvc2}
\end{eqnarray}
 This operator is clearly self-adjoint on $L^{2}\left(\Gamma_{0}\right)$
for all $t$. If we switch back to our original time dependent graph,
the condition above translates into the following magnetic vertex
condition: 
\begin{eqnarray}
 &  & \psi_{e}\left(v,t\right)=\psi_{e'}\left(v,t\right),\ \ \forall e,e'\in S_{v},\label{oggraphvc}\\
 &  & \sum_{e\in S_{v}}\left(\frac{\partial}{\partial x_{e}}-i\cdot\text{sgn}\left(x_{e}\right)\frac{\dot{L}_{e}\left(t/T\right)}{4T}\right)\psi_{e}\left(x_{e},t\right)=0\ \ {\rm at\ }\ |x_{e}|=\frac{L_{e}\left(t/T\right)}{2}.\label{oggraphvc2}
\end{eqnarray}
 For every $t$, our (time dependent) magnetic Laplacian is unitarily
equivalent to a self-adjoint operator on $L^{2}\left(\Gamma_{0}\right)$,
and is thus self-adjoint for all $t$. One can also verify that the
corresponding time evolution is now unitary.

We thus conclude that for the problem of the time dependent graph
to be well posed, the usual Neumann-Kirchhoff vertex condition should
be replaced with the magnetic vertex condition (\ref{oggraphvc}),
(\ref{oggraphvc2}). Moreover, note that in the limit $T\rightarrow\infty$,
one recovers the standard Neumann-Kirchhoff vertex condition of a
stationary quantum graph. We state this as the main result for this
section:
\begin{thm}
\label{theorem-kirchhoff} The Schrödinger equation (\ref{grapheq})
on the stationary metric graph $\Gamma_{0}$ equipped with the magnetic
vertex conditions (\ref{graphvc}, \ref{graphvc2}) induces a unitary
flow on $L^{2}\left(\Gamma_{0}\right)$. This allows to define a solution
to the Schrödinger equation (\ref{laplacian}) on the time dependent
graph $\Gamma$ with unitary time evolution, by composing the solutions
of (\ref{grapheq}) with the unitary family $U_{t}$ (\ref{unitary3},
\ref{unitary4}). 
\end{thm}

This theorem is a particular case of Theorem \ref{thm:general-condition}
in Section \ref{sec:Generalizations}, where the proof can be found.

We now go back to our representation of the problem using the stationary
graph (\ref{graphvc}), (\ref{graphvc2}). It turns out that using
this representation, the problem can be solved more easily.

Note that the magnetic potential of the instantaneous operator is
given on each edge by $\mathcal{A}_{e}\left(\xi_{e}\right)=\xi_{e}\dot{L}_{e}L_{e}/2T$.
Since each edge is parameterized by $\left[-1/2,1/2\right]$ and $\mathcal{A}$
is an anti-symmetric function, then the integral of $\mathcal{A}$
along every edge is equal to zero. In particular, this shows that
the total magnetic flux induced by the magnetic potential through
every cycle of the graph vanishes. Thus, following the method presented
in \cite[sec. 2.6]{BerKuc_graphs}, one can eliminate the magnetic
term by applying the gauge transformation 
\begin{equation}
\omega_{e}\left(\xi_{e},\tau\right)=e^{\frac{i}{2}\theta_{e}(\tau)\left(\xi_{e}^{2}-1/4\right)}g_{e}\left(\xi_{e},\tau\right),\ \ \theta_{e}=\frac{\dot{L}_{e}L_{e}}{2T}.\label{graphgauge}
\end{equation}
 The Schrödinger equation then reads
\begin{align}
 & \frac{i}{T}\frac{\partial g_{e}\left(\xi_{e},\tau\right)}{\partial\tau}=\label{finalgraph}\\
 & -\frac{1}{L_{e}^{2}\left(\tau\right)}\cdot\frac{\partial^{2}g_{e}\left(\xi_{e},\tau\right)}{\partial\xi_{e}^{2}}+\left(\frac{L_{e}\ddot{L}_{e}}{4T^{2}}\xi_{e}^{2}-\frac{\ddot{L}_{e}L_{e}+\dot{L}_{e}^{2}}{16T^{2}}\right)g_{e}\left(\xi_{e},\tau\right)\ ,\ |\xi_{e}|\leq\frac{1}{2},\ \forall e\in\mathcal{E},\nonumber 
\end{align}
 where the vertex conditions are now
\begin{eqnarray}
 &  & L_{e}^{-1/2}\left(\tau\right)g_{e}\left(v,\tau\right)=L_{e'}^{-1/2}\left(\tau\right)g_{e'}\left(v,\tau\right),\ \ \forall e,e'\in S_{v},\label{graphvc3}\\
 &  & \sum_{e\in S_{v}}\frac{1}{L_{e}^{3/2}\left(\tau\right)}\cdot\frac{\partial g_{e}\left(v,\tau\right)}{\partial\xi_{e}}=0.\label{graphvc4}
\end{eqnarray}

In general, the time dependent potential and vertex conditions make
the Schrödinger equation (\ref{finalgraph}) non-separable, and thus
difficult to solve. One way to simplify the equation is by assuming
the adiabatic limit, $T\rightarrow\infty$, corresponding to the edge
lengths changing very slowly compared to the characteristic time scale
of the system. In this limit, the solution to the Schrödinger equation
is given by
\begin{equation}
g_{e}\left(\xi,\tau\right)=e^{-i\int_{0}^{\tau}H(\tau')d\tau'}g_{e}\left(\xi,0\right),\label{eq:propogator}
\end{equation}
where $H\left(\tau\right)$ is the instantaneous operator at time
$\tau$. Meaning -- the time evolution at the adiabatic limit can
be obtained by computing the instantaneous spectrum.

Note that as $T\rightarrow\infty$, the shifted harmonic potential
in (\ref{finalgraph}) is negligible. Namely, the difference between
our instantaneous Schrödinger operator and the weighted Laplacian
without the potential converges in norm to zero. The instantaneous
equation for the eigenfunctions $\{g_{e}\left(\xi_{e},\tau\right)\}_{e\in\mathcal{E}}$
can thus be solved via the standard scattering approach presented
in \cite{KotSmi_prl00}, as described in Appendix \ref{sec:Secular}.

\subsection{Example -- equilateral graph with linearly growing edges\label{subsec:Example-linear}}

We demonstrate the above theory by computing the time evolution for
a very simple system -- an equilateral graph whose edge lengths grow
linearly in time:
\begin{equation}
L_{e}\left(t/T\right):=L\left(t/T\right)=L_{0}+\frac{Vt}{T}.\label{eq:linear-edges}
\end{equation}
To solve this problem, we consider the associated stationary metric
graph, $\Gamma_{0}$. The main feature of this system which simplifies
the computation is the fact that $\ddot{L}_{e}=0$, and so the harmonic
term in the Schrödinger equation (\ref{finalgraph}) vanishes, leaving
us with the Schrödinger equation on $\Gamma_{0}$
\begin{align}
 & \frac{i}{T}\frac{\partial g_{e}\left(\xi_{e},\tau\right)}{\partial\tau}=-\frac{1}{L^{2}\left(\tau\right)}\cdot\frac{\partial^{2}g_{e}\left(\xi_{e},\tau\right)}{\partial\xi_{e}^{2}}-\frac{V^{2}}{16T^{2}}g_{e}\left(\xi_{e},\tau\right),\label{eq:simplified-1}
\end{align}
where the instantaneous operator is simply the Laplacian shifted by
a constant. Moreover, since all edges are of equal length, the boundary
conditions (\ref{graphvc3}, \ref{graphvc4}) are simply the standard
Neumann-Kirchhoff condition.

Since the vertex conditions for this system are time independent,
the given problem is separable. A straightforward computation using
separation of variables gives that the solution of (\ref{eq:simplified-1})
admits the following series expansion on each edge:
\begin{align}
 & g_{e}\left(\xi_{e},\tau\right)=\sum_{n\in\mathbb{N}}^{\infty}a_{n}g_{e}^{\left(n\right)}\left(\xi_{e}\right)e^{iT\lambda_{n}\frac{1}{V\left(L_{0}+V\tau\right)}},\label{eq:eigenstates}
\end{align}
where $\lambda_{n}$ are the eigenvalues of the stationary graph $\Gamma_{0}$
imposed with the Neumann-Kirchhoff condition and $g_{e}^{\left(n\right)}$
are the associated eigenfunctions. It is thus enough to consider the
time evolution for initial conditions of the form $g_{e}^{\left(n\right)}\left(\xi_{e}\right)$.

\begin{figure}
\centerline{ \includegraphics[width=0.55\textwidth]{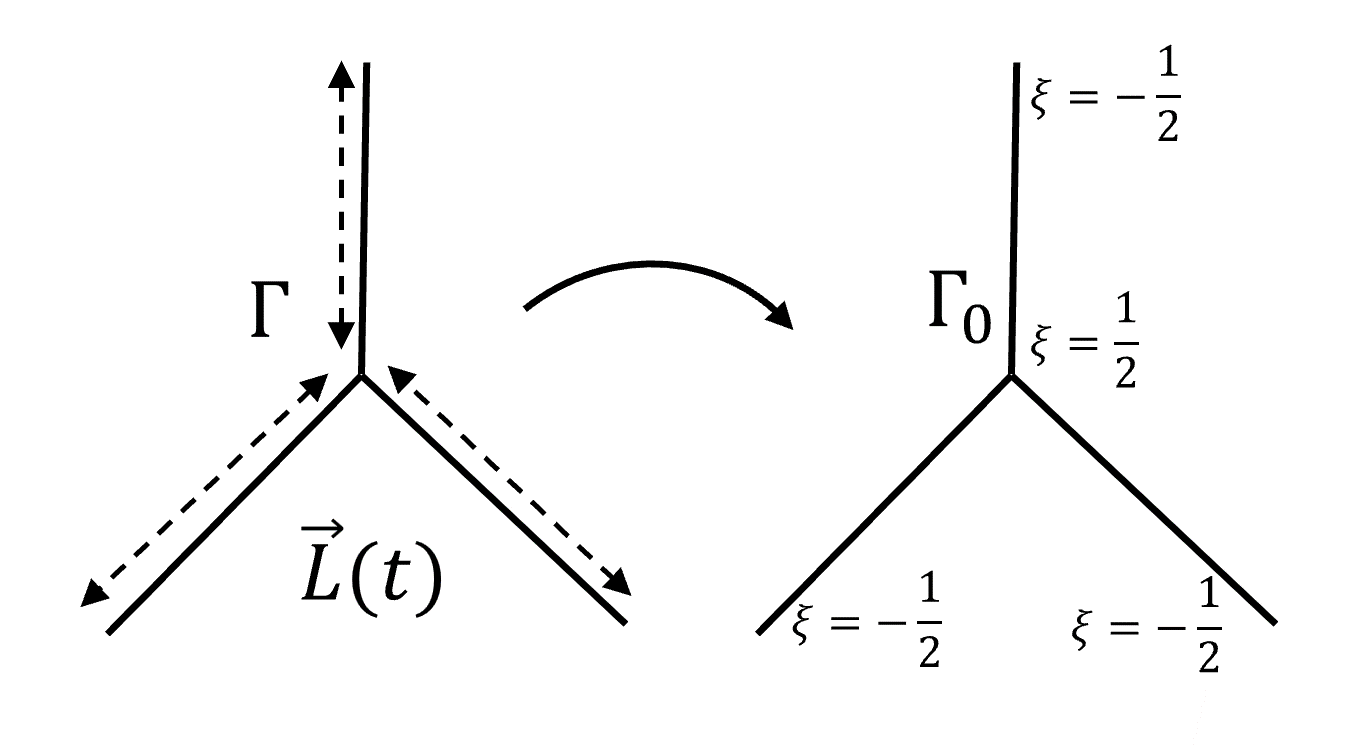} }
\caption{\label{fig:t-star} An equilateral star graph with three edges of
varying lengths $\Gamma_{t}$, along with the associated stationary
graph $\Gamma_{0}$ parametrized so that the central vertex corresponds
to $\xi=1/2$.}
\end{figure}

As a more concrete example, we consider the equilateral star graph
with $E$ edges (see Figure \ref{fig:t-star}), parameterized so that
central vertex corresponds to $\xi_{e}=1/2$. The eigenvalues for
this graph can be written explicitly as $\left\{ \lambda_{n}\right\} _{n=0}^{\infty}\cup\left\{ \alpha_{n}\right\} _{n=0}^{\infty}$,
where
\begin{align*}
 & \lambda_{n}=\left(\pi n\right)^{2}-\frac{V^{2}}{16T^{2}},\text{ (Multiplicity \ensuremath{1})},\\
 & \alpha_{n}=\left(\frac{\pi}{2}+\pi n\right)^{2}-\frac{V^{2}}{16T^{2}},\text{ (Multiplicity \ensuremath{E-1}).}
\end{align*}
The associated normalized eigenfunctions are given by
\begin{align*}
 & g_{e}^{\left(\lambda_{0}\right)}\left(\xi_{e}\right)=\frac{1}{\sqrt{E}},\\
 & g_{e}^{\left(\lambda_{n}\right)}\left(\xi_{e}\right)=\sqrt{\frac{2}{E}}\cos\left(n\pi\left(\xi+\frac{1}{2}\right)\right)\text{ for }n>0,\\
 & g_{e}^{\left(\alpha_{n}\right)}\left(\xi_{e}\right)=\sqrt{\frac{2}{E}}\sigma_{e}^{\left(n\right)}\cos\left(\left(\frac{\pi}{2}+\pi n\right)\left(\xi+\frac{1}{2}\right)\right),
\end{align*}
where $\sigma_{e}^{\left(n\right)}\in\left\{ -1,0,1\right\} $ are
chosen on each edge to give $E-1$ linearly independent functions
satisfying the vertex condition at the central vertex.

For convenience, we focus on the time evolution for the eigenstates
$g_{e}^{\left(\lambda_{n}\right)}$ with $n>0$. We can compute the
time evolution for these states on the time dependent star graph $\Gamma$
by applying the inverse gauge transformation to (\ref{graphgauge})
and unitary family $U_{t}^{-1}$:
\begin{align}
 & \psi_{e}^{\left(\lambda_{n}\right)}\left(x_{e},t\right)=\label{eq:linear-solution}\\
 & \sqrt{\frac{2}{EL\left(t/T\right)}}\cos\left(n\pi\left(\frac{x}{L\left(t/T\right)}+\frac{1}{2}\right)\right)e^{i\frac{\pi^{2}n^{2}T-V^{2}/16T}{VL\left(t/T\right)}}e^{i\frac{V}{4T}\left(\frac{x^{2}}{L\left(t/T\right)}-\frac{L\left(t/T\right)}{4}\right)}.\nonumber 
\end{align}
One can easily verify that $\psi_{e}^{\left(\lambda_{n}\right)}$
indeed solves the Schrödinger equation (\ref{laplacian}), satisfies
the magnetic boundary conditions (\ref{oggraphvc}, \ref{oggraphvc2}),
and that $\left\Vert \Psi^{\left(\lambda_{n}\right)}\left(x,t\right)\right\Vert _{L^{2}}=1$
for all $t$. By writing the initial condition as a generalized Fourier
series in the functions $g_{e}^{\left(\lambda_{n}\right)},g_{e}^{\left(\alpha_{n}\right)}$
on each edge, one may compute the time evolution of (\ref{laplacian})
for any initial condition.

Using (\ref{eq:linear-solution}), one can for instance compute the
average kinetic energy of the states $\Psi^{\left(\lambda_{n}\right)}$
over time, as done in \cite{Matrasulov2015}:
\begin{align}
 & \left\langle K\left(t\right)\right\rangle _{\Psi^{\left(n\right)}}:=\left\langle \Psi^{\left(\lambda_{n}\right)}|H|\Psi^{\left(\lambda_{n}\right)}\right\rangle \left(t\right)=\sum_{e\in\mathcal{E}}\int_{-L_{e}/2}^{L_{e}/2}\left|\frac{\partial\psi_{e}^{\left(\lambda_{n}\right)}\left(x,t\right)}{\partial x_{e}}\right|^{2}dx\label{eq:mean-kinetic}\\
 & =\frac{2}{EL^{3}}\sum_{e\in\mathcal{E}}\int_{-L_{e}/2}^{L_{e}/2}\left(\frac{x^{2}\dot{L}^{2}}{4T^{2}}\cos^{2}\left(n\pi\left(\frac{x}{L}+\frac{1}{2}\right)\right)+\pi^{2}n^{2}\sin^{2}\left(n\pi\left(\frac{x}{L}+\frac{1}{2}\right)\right)\right)dx\nonumber \\
 & =\frac{V^{2}}{2T^{2}}\left(\frac{1}{4\pi^{2}n^{2}}+\frac{1}{24}\right)+\frac{\pi^{2}n^{2}}{\left(L_{0}+Vt/T\right)^{2}},\,\,\,\,\,\left(n>0\right).\nonumber 
\end{align}
Namely, the average kinetic energy of the particle decreases over
time, and approaches a constant which depends entirely on the growth
rate of the edges, $V$. One can show that the same result holds for
all eigenstates, except for $g_{e}^{\left(0\right)}$, whose average
kinetic energy remains constant in time.

\section{Generalizations and applications\label{sec:Generalizations}}

\subsection{Generalization of the results to time dependent vertex conditions}

The analysis above focused on the standard Neumann-Kirchhoff vertex
conditions. Nevertheless, one would generally like to consider various
vertex conditions, such as the Dirichlet condition (considered in
\cite{Matrasulov2015}), or the related $\delta$ and $\delta'$ vertex
conditions (see \cite[sec. 1.4.4]{BerKuc_graphs}). We have seen in
Section \ref{section1} that in the case of the Dirichlet condition,
there is no need to reformulate the problem so that it is well posed.
Yet, for an arbitrary self-adjoint vertex condition which is valid
for a stationary metric graph, one would need to make the appropriate
adjustments when considering the moving quantum graph (as in the case
of the Neumann-Kirchhoff condition).

In order to present these adjustments, we first give a brief reminder
about how general self-adjoint vertex conditions are written for a
stationary quantum graph. A more in-depth survey of the topic can
be found in \cite[sec. 1.4.1]{BerKuc_graphs}. Given a Sobolev function
$f\in H^{2}\left(\Gamma\right)$ and a vertex $v\in\mathcal{V}$,
we denote by $F\left(v\right)\in\mathbb{C}^{d_{v}}$ the vector of
boundary values of $f$ at the vertex $v$ along the edges adjacent
to $v$:
\begin{equation}
F\left(v\right):=\left(f_{1}\left(v\right),...,f_{d_{v}}\left(v\right)\right)^{t}.\label{eq:f-vec}
\end{equation}
 Similarly, we denote the vector of (outwards pointing) derivatives
of $f$ at $v$ by $F'\left(v\right)$.
\begin{thm}
\label{thm:-kuchment}\cite{KosSch_jpa99,BerKuc_graphs} Let $\Gamma$
be a compact metric graph. Consider the operator $\Delta=-\frac{d^{2}}{dx^{2}}$
with domain consisting of all functions $f\in H^{2}\left(\Gamma\right)$
satisfying certain local conditions at each vertex. Then $\Delta$
is self-adjoint if and only if for every vertex $v$ of degree $d_{v}$
there exist $d_{v}\times d_{v}$ matrices $A_{v}$ and $B_{v}$ such
that

1. The $d_{v}\times2d_{v}$ matrix $\left(A_{v}|B_{v}\right)$ is
of maximal rank,

2. The matrix $A_{v}B_{v}^{*}$ is self-adjoint,

3. Each $f\in Dom\left(\Delta\right)$ satisfies
\begin{equation}
A_{v}F\left(v\right)+B_{v}F'\left(v\right)=0.\label{eq:kuchment-mat}
\end{equation}
\end{thm}

In other words, the possible vertex conditions of a stationary quantum
graph are parameterized by the pair $\left(A_{v},B_{v}\right)$ at
each vertex $v$.

In general, it would be natural to associate with the moving graph
$\Gamma$ and each $v\in\mathcal{V}$ a continuous family $\left(A_{v}\left(t\right),B_{v}\left(t\right)\right)_{t\in\left[0,T\right]}$
which satisfies conditions $1-3$ above. This would give a family
of time varying vertex conditions associated with the time dependent
metric graph. Yet, this does not in general provide unitary time evolution
(as in the case of the Neumann-Kirchhoff condition). This problem
arises once again due to the fact that the changing edges of the graph
induce a magnetic potential as seen in Equation (\ref{finalgraph}),
and the associated pair $\left(A_{v}\left(t\right),B_{v}\left(t\right)\right)$
might not render the given magnetic operator self-adjoint. To solve
this, we employ the following standard result:
\begin{thm}
\label{thm:magneticflux}\cite[thm 2.6.1]{BerKuc_graphs} Let $\Gamma$
be a quantum graph with a magnetic potential $\mathcal{A}$. Assume
that the magnetic flux of $\mathcal{A}$ through $\Gamma$ vanishes.
Then for any pair $\left(A_{v},B_{v}\right)$ as in theorem \ref{thm:-kuchment},
the magnetic Laplacian $-\left(\frac{\partial}{\partial x}-i\mathcal{A}\left(x\right)\right)^{2}$
is self-adjoint when the graph is endowed with the vertex conditions
\begin{equation}
A_{v}F\left(v\right)+B_{v}DF\left(v\right)=0,\label{eq:AB-condition}
\end{equation}
where $DF\left(v\right)$ is the vector of magnetic derivatives of
$f$ at $v$:
\begin{equation}
Df_{e}\left(x\right)=\left(\frac{\partial}{\partial x_{e}}-i\mathcal{A}_{e}\left(x_{e}\right)\right)f_{e}\left(x_{e}\right).\label{eq:magnetic-deriv}
\end{equation}
\end{thm}

We can combine the two results above with the derivation in Section
\ref{section2}. Motivated by our family of unitaries $U_{t}$, we
define the modified boundary vectors $W\left(v\right),W'\left(v\right)\in\mathbb{C}^{d_{v}}$:
\begin{align}
 & W\left(v\right):=\left(L_{1}^{1/2}f_{1}\left(v\right),...,L_{d_{v}}^{1/2}f_{d_{v}}\left(v\right)\right)^{t},\label{eq:W}\\
 & W'\left(v\right):=\left(L_{1}^{3/2}f_{1}'\left(v\right),...,L_{d_{v}}^{3/2}f_{d_{v}}'\left(v\right)\right)^{t}.\label{eq:DW}
\end{align}
In this new representation, the condition $A_{v}F+B_{v}F'$ translates
into
\begin{align}
 & A_{v}L_{v}^{-1/2}W\left(v\right)+B_{v}L_{v}^{-3/2}W'\left(v\right)=0,\label{eq:W-cond}\\
 & L_{v}:=\text{diag}\left(L_{1},...,L_{d_{v}}\right).\label{eq:L-diag}
\end{align}

Then combining the results above immediately provides the following:
\begin{prop}
\label{prop:self-adjoint}Let $\Gamma$ be a time-dependent metric
graph. Let $\left(A_{v}\left(t/T\right),B_{v}\left(t/T\right)\right)_{t\in\left[0,T\right]}^{v\in\mathcal{V}}$
be a given family of pairs of matrices which satisfy that for all
$t,v$:

1. The $d_{v}\times2d_{v}$ matrix $\left(A_{v}\left(t/T\right)|B_{v}\left(t/T\right)\right)$
has maximal rank, 

2. The matrix $A_{v}B_{v}^{*}\left(t/T\right)$ is self-adjoint.

Then for all $\tau:=\frac{t}{T}\in\left[0,1\right]$, the following
operator on $L^{2}\left(\Gamma_{0}\right)$ is self-adjoint:
\begin{align}
 & H_{\tau}|_{e}\omega_{e}=\frac{1}{L_{e}\left(\tau\right)^{2}}\left[-\left(\frac{\partial}{\partial\xi_{e}}-i\xi_{e}\frac{\dot{L}_{e}\left(\tau\right)L_{e}\left(\tau\right)}{2T}\right)^{2}-\xi_{e}^{2}\left(\frac{\dot{L}_{e}\left(\tau\right)L_{e}\left(\tau\right)}{2T}\right)^{2}\right]\omega_{e},\label{eq:3.3-schrod}\\
 & Dom\left(H_{\tau}\right)=\left\{ \omega\in H^{2}\left(\Gamma_{0}\right):A_{v}\left(\tau\right)L_{v}^{-1/2}W\left(v\right)+B_{v}\left(\tau\right)L_{v}^{-3/2}D_{\tau}W\left(v\right)=0\right\} ,
\end{align}
where
\begin{align}
 & W\left(v\right):=\left(\omega_{1}\left(v\right),...,\omega_{d_{v}}\left(v\right)\right)^{t},\label{eq:omega-def}\\
 & D_{\tau}W\left(v\right):=\left(D_{\tau}\omega_{1}\left(v\right),...,D_{\tau}\omega_{d_{v}}\left(v\right)\right)^{t},
\end{align}
 and $D_{\tau}$ is the magnetic derivative:
\begin{equation}
D_{\tau}\omega_{e}\left(\xi_{e}\right):=\left(\frac{\partial}{\partial\xi_{e}}-i\xi_{e}\frac{\dot{L}_{e}L_{e}\left(\tau\right)}{2T}\right)\omega_{e}\left(\xi_{e}\right).\label{eq:3.3-magnetic-deriv}
\end{equation}
\end{prop}

We are now in position to prove the following generalization of Theorem
\ref{theorem-kirchhoff}:
\begin{thm}
\label{thm:general-condition} Let $\Gamma$ be a time-dependent quantum
graph, along with a family of matrices $\left(A_{v}\left(t/T\right),B_{v}\left(t/T\right)\right)_{t\in\left[0,T\right]}^{v\in\mathcal{V}}$
as in Proposition \ref{prop:self-adjoint}. Assume that

1. The map $t\mapsto\left(A_{v}\left(t/T\right),B_{v}\left(t/T\right)\right)$
is $C^{2}$ for all $t,v$,

2. For each $v\in\mathcal{V}$, the kernel of $B_{v}\left(t/T\right)$
is independent of $t$. 

Then with the operator $H_{\tau}$ from Proposition \ref{prop:self-adjoint},
the Schrödinger equation on $\Gamma_{0}$
\begin{equation}
i\frac{\partial g}{\partial\tau}=H_{\tau}g\label{eq:general-schrod}
\end{equation}
induces a unitary flow on $L^{2}\left(\Gamma_{0}\right)$, which allows
to define via the unitary family $U_{t}$ in (\ref{unitary3}, \ref{unitary4})
a solution to the Schrödinger equation on $\Gamma$:
\begin{align}
 & i\frac{\partial\psi}{\partial t}=-\frac{\partial^{2}\psi}{\partial x^{2}},\label{eq:og-schrodinger}\\
 & A_{v}\left(t/T\right)\Psi\left(v\right)+B_{v}\left(t/T\right)D_{t}\Psi\left(v\right)=0,\,\,\forall v\in\mathcal{V},
\end{align}
where
\begin{equation}
D_{t}\psi_{e}\left(x_{e}\right):=\left(\frac{\partial}{\partial x_{e}}-ix_{e}\frac{\dot{L}_{e}}{2TL_{e}}\right)\psi_{e}\left(x_{e}\right).\label{eq:megnatic-deriv-4}
\end{equation}
\end{thm}

\begin{figure}
\centerline{ \includegraphics[scale=0.75]{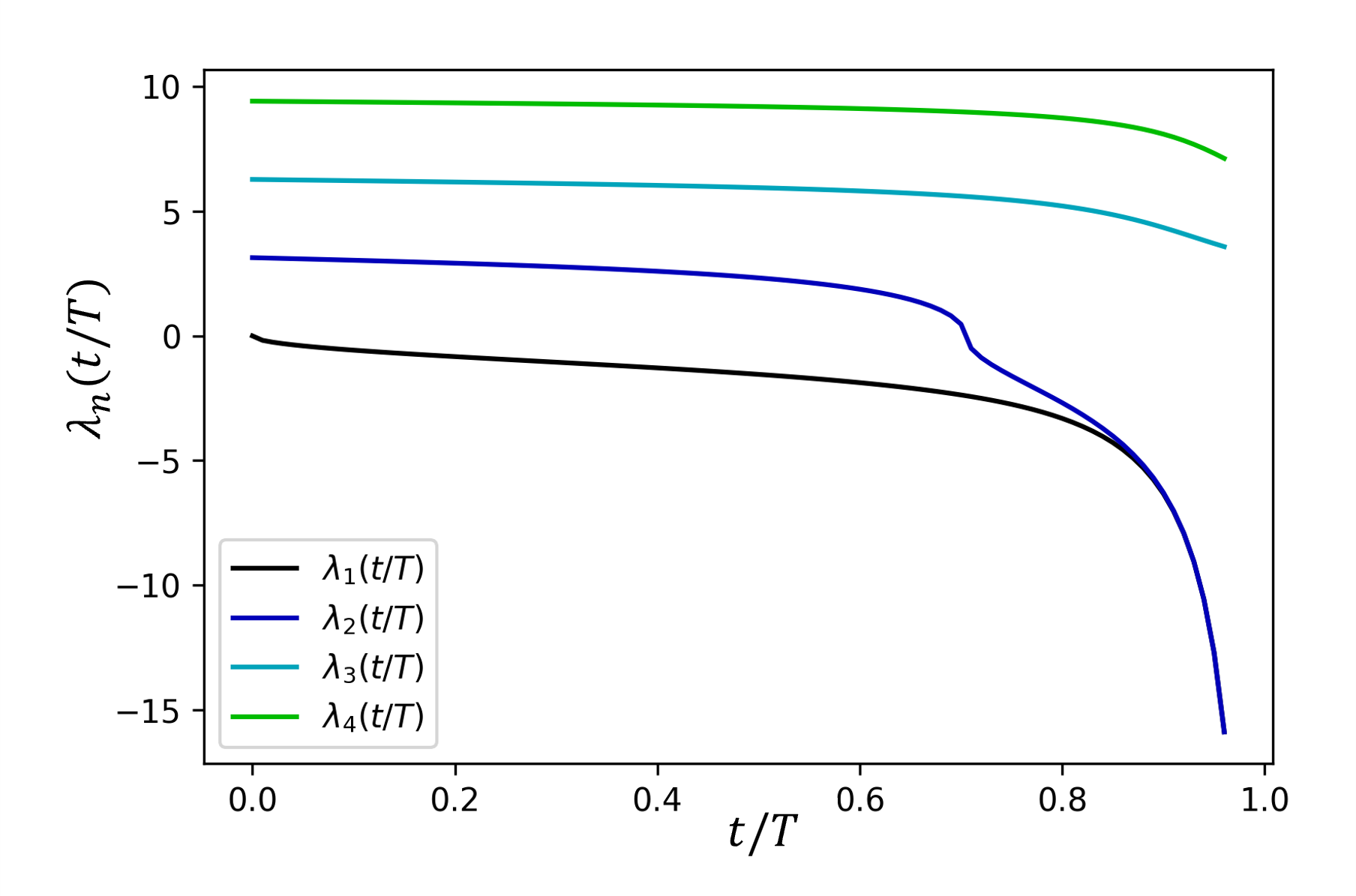} } \caption{\label{fig:Robin-curves} The first four eigenvalues of the Robin
interval as a function of $t/T$. As $t/T\rightarrow1$, two of the
eigenvalue curves approach $-\infty$.}
\end{figure}

\begin{proof}
The argument is similar to the one presented in \cite{Duca2021},
where additional details may be found. We outline the idea of the
proof, focusing on the relevant changes that should be made for our
case. We apply theorem $8.1$ in \cite{Kisynski1964}. By \cite[thm 1.4.11]{BerKuc_graphs},
the quadratic form associated with $H_{\tau}$ is given by
\begin{align}
Q_{\tau}\left[\omega\right] & =\sum_{e\in\mathcal{E}}\int_{-1/2}^{1/2}\frac{1}{L_{e}\left(\tau\right)}\left|\frac{\partial\omega_{e}}{\partial\xi_{e}}-i\xi_{e}\frac{\dot{L}_{e}\left(\tau\right)L_{e}\left(\tau\right)}{2T}\omega_{e}\right|^{2}d\xi_{e}\label{eq:quadratic}\\
 & -\sum_{e\in\mathcal{E}}\int_{-1/2}^{1/2}\left(\frac{\dot{L}_{e}\left(\tau\right)L_{e}\left(\tau\right)}{2T}\right)^{2}\xi_{e}^{2}\left|\omega_{e}\right|^{2}d\xi_{e}\nonumber \\
 & +\sum_{v\in\mathcal{V}}\left\langle \Lambda_{v}P_{R,v}L_{v}^{1/2}W\left(v\right),P_{R,v}L_{v}^{1/2}W\left(v\right)\right\rangle ,\nonumber 
\end{align}
with domain
\begin{equation}
Dom\left(Q_{\tau}\right)=\left\{ \omega\in H^{1}\left(\Gamma_{0}\right)|P_{D,v}L_{v}^{1/2}W\left(v\right)=0\right\} ,\label{eq:domain}
\end{equation}
where $P_{D,v}$ is the orthogonal projection onto $\ker\left(B_{v}\left(\tau\right)\right)$,
and $P_{R,v},\Lambda_{v}$ are matrices which depend smoothly on $A\left(\tau\right),B\left(\tau\right)$.

Due to Proposition \ref{prop:self-adjoint}, the operator $H_{\tau}$
is self-adjoint for all $\tau$. Combining condition ($1$) with Equation
(\ref{eq:quadratic}) yields that the map $\tau\mapsto Q_{\tau}\left[f\right]$
is $C^{2}$ for all $f$. Furthermore, since $P_{D,v}$ is the orthogonal
projection onto $\ker\left(B_{v}\right)$, assumption ($2$) ensures
that the domain of $Q_{\tau}$ is independent of $\tau$. Lastly,
a straightforward computation shows that
\begin{equation}
Q_{\tau}\left[\omega\right]\geq b\left\Vert \omega\right\Vert _{H^{1}\left(\Gamma_{0}\right)}^{2}-c\left\Vert \omega\right\Vert _{L^{2}\left(\Gamma_{0}\right)}^{2},\label{eq:bdd-below}
\end{equation}
for some $b>0,c\in\mathbb{R}$. Under these conditions, \cite{Kisynski1964}
shows that the Schrödinger equation (\ref{eq:general-schrod}) induces
a unitary flow on $L^{2}\left(\Gamma_{0}\right)$. Composing this
flow with the unitary family $U_{t}^{-1}$ induces a unitary flow
on $L^{2}\left(\Gamma\right)$, which due to the computations above,
provides a solution to the Schrödinger equation (\ref{eq:og-schrodinger}). 
\end{proof}
Note that one may employ the same gauge transformation as in Section
\ref{section2} to bring the Schrödinger equation to the simpler form
(\ref{finalgraph}), along with the non-magnetic boundary conditions
\begin{equation}
A_{v}\left(\tau\right)L_{v}^{-1/2}W\left(v\right)+B_{v}\left(\tau\right)L_{v}^{-3/2}W'\left(v\right)=0,\,\,\forall v\in\mathcal{V}.\label{eq:bdry-gauge}
\end{equation}
As before, in the adiabatic limit, this equation may be solved spectrally,
by applying the scattering formalism with the appropriate scattering
matrix.
\begin{rem}
We would like to illustrate the possible importance of the condition
regarding $\ker\left(B\right)$ in Theorem \ref{thm:general-condition}.
To do so, we consider the stationary quantum graph given by the interval
$\left[0,1\right]$, equipped with the time dependent Robin condition
at both endpoints $v\in\left\{ 0,1\right\} $:
\begin{equation}
\sin\left(\frac{\pi}{2T}t\right)f\left(v\right)=-\cos\left(\frac{\pi}{2T}t\right)f'\left(v\right),t\in\left[0,T\right].\label{eq:Robin}
\end{equation}
We focus on the adiabatic case, where the boundary conditions are
slowly varied from the Neumann condition at $t=0$ to the Dirichlet
condition at $t=T$. While the associated family $\left(A\left(t/T\right),B\left(t/T\right)\right)$
is analytic, it is evident that $\ker\left(B\left(t/T\right)\right)$
changes at $t=T$. Consequently, one can show that the first two eigenvalues
$\lambda_{1}\left(t\right),\lambda_{2}\left(t\right)$ approach the
value $-\infty$ as $t\rightarrow T$ (see Figure \ref{fig:Robin-curves}).
While we cannot explicitly compute the time evolution for this system
in the non-adiabatic case, this unusual spectral behavior hints that
the time evolution for the associated states may display an abrupt
change of behavior for $t>T$.

The phenomenon of eigenvalues approaching $-\infty$ as $\ker\left(B\left(t/T\right)\right)$
changes has appeared from different perspectives in several recent
and ongoing works. This phenomenon can be identified not only through
$\ker\left(B\left(t/T\right)\right)$, but also using quantities such
as the spectral flow (see \cite{Sofer2022,Band}) and the Duistermaat
index, as shown by Berkolaiko, Cox, Latushkin and Sukhtaiev \cite{berkolaiko2023duistermaat}.
\end{rem}

\subsection{Application -- the geometric phase}

We apply the theory above to show the existence of a geometric phase
for slowly varying quantum graphs, similar to the one proposed by
Michael Berry in \cite{Berry1984}. There, Berry considered time evolution
of the Schrödinger equation in systems with a time dependent Hamiltonian
$H\left(\boldsymbol{R}(t)\right)$, where the time dependent parameter
$\boldsymbol{R}\left(t\right)\in\mathbb{R}^{n}$ is assumed to change
adiabatically. He then showed that if the system is initially prepared
at an eigenstate $\psi_{n}\left(x,0\right)$ of $H\left(\boldsymbol{R}(0)\right)$,
then at the adiabatic limit, the time evolution is given by
\begin{equation}
\psi\left(x,t\right)=\exp\left(-\frac{i}{\hbar}\int_{0}^{t}E_{n}\left(\boldsymbol{R}(t')\right)dt'\right)\exp\left(i\gamma_{n}\left(t\right)\right)\psi_{n}\left(x,0\right),\label{eq:berryphase}
\end{equation}
where $E_{n}\left(\boldsymbol{R}(t)\right)$ is the $n$th eigenvalue
of $H\left(\boldsymbol{R}(t)\right)$. The first exponential term
is known as the dynamical phase, and it is a direct analog of the
phase acquired by solutions of the Schrödinger equation in time independent
systems. The second exponential term, known as the geometric phase
(or Berry phase), is given by
\begin{equation}
\gamma_{n}\left(t\right)=i\int_{0}^{t}\left\langle \psi_{n}\left(\boldsymbol{R}(t')\right)|\nabla_{\boldsymbol{R}}\psi_{n}\left(\boldsymbol{R}(t')\right)\right\rangle \cdot\dot{\boldsymbol{R}}\left(t'\right)dt',\label{eq:bphase2}
\end{equation}
where $\psi_{n}\left(\boldsymbol{R}(t)\right)$ is the eigenfunction
associated with $E_{n}\left(\boldsymbol{R}(t)\right)$.

Since its discovery in $1982$, the geometric phase has been the subject
of many works, and has been experimentally measured in various settings.
In \cite{Simon1983}, Simon showed that if the adiabatic time evolution
of the eigenstate is considered as parallel transport in the associated
space of parameters, then the geometric phase can be thought of as
the corresponding holonomy.

The original derivation by Berry was given for systems with time dependent
Hamiltonians on a fixed Hilbert space. This does not exactly fit the
time dependence in our setting, where the geometry itself of the system
changes in time, causing the Hilbert space to change. Nevertheless,
Theorem \ref{thm:general-condition} shows that the time evolution
for our system is equivalent to that of a stationary metric graph
with a time dependent Hamiltonian. Utilizing this equivalence, one
can repeat the original derivation by Berry to conclude that (\ref{eq:berryphase})
holds for time dependent quantum graphs as well. Namely, for the Schrödinger
equation (\ref{eq:og-schrodinger}) on a graph with a time dependent
vector of edge lengths $\mathcal{L}$, we have the adiabatic result
as $T\rightarrow\infty$:
\begin{equation}
\gamma_{n}\left(t\right)=i\int_{0}^{t}\left\langle \psi_{n}\left(\mathcal{L}(t')\right)|\nabla_{\mathcal{L}}\psi_{n}\left(\mathcal{L}(t')\right)\right\rangle \cdot\dot{\mathcal{L}}\left(t'\right)dt'.\label{eq:berrylengths}
\end{equation}

Note that differentiation of the eigenvalues and eigenfunctions with
respect to the lengths is possible due to \cite{BerKuc_incol12},
under the assumption that the spectrum of $H\left(\mathcal{L}\left(t\right)\right)$
is simple along the entire path. In general, the time dependence of
the system can also be manifested by parametrically changing the boundary
conditions in time (as in Theorem \ref{thm:general-condition}).

Since the potential appearing in (\ref{finalgraph}) is real valued,
our Hamiltonian possesses time reversal symmetry, and the eigenfunctions
of $H\left(\mathcal{L}(t)\right)$ all remain real when parallel transported
along the path $\mathcal{L}\left(t\right)$. Thus, by \cite{Simon1983},
the geometric phase must be an integer multiple of $\pi$ when the
trajectory in parameter space $\mathcal{L}\left(t\right)$ is taken
to be closed. For instance, this happens if the graph $\Gamma$ changes
periodically, i.e. $\mathcal{L}\left(t+T\right)=\mathcal{L}\left(t\right)$
for some $T>0$.

Let us consider the simplest nontrivial example, of a star graph with
three edges of varying length, where the system is initially prepared
at the first non-constant eigenstate, $\psi_{2}$. The parameter space
in this case is $\mathbb{R}_{+}^{3}$, with the diagonal line $\ell:=\text{span}\left\{ \left(1,1,1\right)\right\} $
removed (since the eigenstate $\psi_{2}$ is degenerate along this
line). Due to the invariance of the geometric phase under homotopy,
it is enough to compute the geometric phase along a single cycle which
winds once around $\ell$. For instance, we can choose some circle
around the line $\ell$ which is contained in the plane $L_{1}+L_{2}+L_{3}=1$.
Conveniently, due to the constraint $L_{1}+L_{2}+L_{3}=1$, the Hamiltonian
$H$ can be thought of as depending on only two parameters along the
path (say, $L_{1}$ and $L_{2}$). Our cycle then encloses the isolated
conical point $\left(\frac{1}{3},\frac{1}{3},\frac{1}{3}\right)$
on our plane. We may then apply the result proven in \cite{MR3831157}:
\begin{thm}
\label{thm:conical-point} Let $H\left(\vec{L}\right)$ be a self-adjoint
operator which depends analytically on two parameters $\vec{L}=\left(L_{1},L_{2}\right)$,
with a non-degenerate conical point at some point $\vec{L}_{0}$.
Assume that $H\left(\vec{L}\right)$ commutes with some anti-unitary
involution $V$. i.e.
\begin{equation}
V\left(\alpha f\right)=\overline{\alpha}V\left(f\right),\,\,V^{2}=1,\,\,\left\langle Vf,Vg\right\rangle =\left\langle f,g\right\rangle .\label{eq:involution}
\end{equation}
 Then the geometric phase along a cycle enclosing the singularity
$\vec{L}_{0}$ is $\pi$ modulo $2\pi$.
\end{thm}

By applying the theorem above on $H_{\tau}$ along with the anti-unitary
involution $V\psi\left(x\right)=\overline{\psi\left(x\right)}$, we
obtain the following:
\begin{cor}
The geometric phase acquired by $\psi_{2}$ along a closed cycle $\mathcal{L}\left(t\right)$
is equal to $\pi$ times the winding number of $\mathcal{L}\left(t\right)$
around the degenerate line $\ell$.
\end{cor}

More generally, for any eigenstate $\psi_{n}$ of the star graph there
exists a finite collection $\left\{ \ell_{j}\right\} _{j=1}^{N}$
of lines through the origin along which $\psi_{n}$ is degenerate.
In this case, the same argument shows that the geometric phase will
be equal to $\pi$ times the sum of winding numbers of the path around
the given lines. For an arbitrary graph, the computation of the geometric
phase is more complicated, since the subset $D$ of the parameter
space $\mathbb{R}_{+}^{\left|\mathcal{E}\right|}$ such that $\psi_{n}$
is degenerate might be substantially harder to compute (by Wigner
and Von-Neumann, it is generically a subset of codimension two). As
before, the possible values for the geometric phase will be determined
by the topology of the space $\mathbb{R}_{+}^{\left|\mathcal{E}\right|}-D$.

We shall mention that an equivalent computation of the geometric phase
can be obtained by replacing the parameter space $\mathbb{R}_{+}^{\left|\mathcal{E}\right|}$
with a different one, known as the \emph{secular manifold}. The secular
manifold can be thought of as a quotient of our original parameter
space, which takes into account the invariance of our system under
scaling of the edges (For a thorough introduction to the secular manifold,
see \cite{AloBanBer_cmp18,Alon_PhDThesis}). Such a computation for
a star graph was recently done by Lior Alon \cite{Alona} independently
of the authors. The advantage of this approach is that it reduces
the computation of the geometric phase to that of a single monodromy
for all eigenstates $\psi_{n}$. The drawback of this approach is
that the given path that the edge lengths complete in parameter space
is harder to express.
\begin{acknowledgement*}
We are grateful to Professor Michael Berry for explaining the physical
origin of the magnetic phase. Thanks are due to Professor Kirill Cherednichenko
for his continuous interest, and for pointing out the similarity between
the boundary conditions in this work and the ones corresponding to
graphs consisting of composite conductors. We also thank Lior Alon
for very helpful discussions which helped understanding the geometric
phase, and to Gregory Berkolaiko for extremely helpful remarks which
corrected and improved this work. We also appreciate the important
and useful comments made by the referees. US thanks the University
of Bath for the hospitality while holding a David Parkin Professorship
(September 2021- March 2022), and GS thanks the Weizmann Institute
for the hospitality while visiting during 2022-2023.
\end{acknowledgement*}

\appendix

\section{\label{sec:Secular} Secular equation for the instantaneous graph}

We now derive the secular equation which provides the eigenvalues
and eigenfunctions of the instantaneous operator, and in turn allows
to solve for the time evolution at the adiabatic limit.

To solve for the instantaneous spectrum, we write our eigenfunctions
on each edge as
\begin{equation}
g_{e}\left(\xi_{e},\tau\right)=a_{e}(\tau)e^{ikL_{e}(\tau)\left(1/2+\xi_{e}\right)}+a_{\hat{e}}(\tau)e^{ikL_{e}(\tau)(1/2-\xi_{e})}.
\end{equation}
Using the modified vertex conditions (\ref{graphvc3}, \ref{graphvc4}),
straightforward algebraic manipulation (see \cite{Berkolaiko_qg-intro17}
for further details) then shows that the scattering coefficients satisfy
\begin{equation}
a_{e}(\tau)=\frac{2L_{e}^{1/2}(\tau)}{d_{v}}\sum_{j\in S_{v}}L_{j}^{-1/2}(\tau)e^{ikL_{j}\left(\tau\right)}a_{\hat{j}}(\tau)-e^{ikL_{e}\left(\tau\right)}a_{\hat{e}}(\tau).\label{eq:linear-0}
\end{equation}

We can equivalently write this in terms of the bond scattering matrix.
We say that the directed edge $j'$ follows the directed edge $j$
if the starting vertex of $j'$ is the end vertex of $j$. With this,
we define the instantaneous bond scattering matrix as
\[
S_{j'j}\left(\tau\right):=\begin{cases}
\frac{2}{d_{v}}\left(\frac{L_{j'}(\tau)}{L_{j}(\tau)}\right)^{1/2}-1, & j'=\hat{j},\\
\frac{2}{d_{v}}\left(\frac{L_{j'}(\tau)}{L_{j}(\tau)}\right)^{1/2}, & j'\text{ follows }j\text{ and }j'\neq\hat{j},\\
0, & \text{otherwise}.
\end{cases}
\]
The relation (\ref{eq:linear-0}) then translates into
\begin{equation}
\left(I-S\left(\tau\right)e^{\rmi k\L\left(\tau\right)}\right)\vec{a}=0,\label{eq:-2}
\end{equation}
 where $\boldsymbol{L}\left(\tau\right):=\mathrm{diag}\left(L_{1}\left(\tau\right),L_{1}\left(\tau\right),L_{2}\left(\tau\right),L_{2}\left(\tau\right),...,L_{\left|\mathcal{E}\right|}\left(\tau\right),L_{\left|\mathcal{E}\right|}\left(\tau\right)\right)$
is the diagonal matrix of edge lengths and $\vec{a}:=\left(a_{1},a_{\hat{1}},...,a_{\left|\mathcal{E}\right|},a_{\hat{\left|\mathcal{E}\right|}}\right)$
is the vector of scattering coefficients.

We now see that $k^{2}>0$ is an eigenvalue of the instantaneous problem
if and only if it is a root of the secular equation: 
\begin{equation}
\det\left(I-S\left(\tau\right)e^{ik\boldsymbol{L}\left(\tau\right)}\right)=0.\label{secequation}
\end{equation}
Thus, one may obtain the instantaneous spectrum in the adiabatic limit
by solving the instantaneous secular equation for each $\tau$. The
coefficients for the associated eigenfunctions may be then solved
from Equation (\ref{eq:-2}).

Note that unlike many standard cases, here the bond scattering matrix
$S\left(\tau\right)$ is not unitary, since the edges of the graph
were scaled, giving a Laplacian with different (length dependent)
weights on each edge. By defining the matrix $P_{j'j}\left(\tau\right)=\frac{1}{\sqrt{L_{j}\left(\tau\right)}}\delta_{j'j}$,
one can see that $S\left(\tau\right)=P^{-1}\tilde{S}P$, where $\tilde{S}$
is the \emph{unitary} bond scattering matrix of the time dependent
metric graph $\Gamma$ at time $\tau$. Thus, although $S$ is not
unitary, it is similar to the unitary bond scattering matrix of the
non-weighted Neumann-Kirchhoff Laplacian, which keeps the secular
equation invariant.

\bibliographystyle{plain}
\bibliography{GlobalBib_210709}

\end{document}